\documentclass[11pt,preprint]{aastex}
\usepackage{epsfig}
\usepackage{lineno}
\usepackage{epsfig} 
\usepackage{amsmath}
\usepackage{amssymb}
\usepackage{natbib}
\usepackage{graphicx}
\usepackage{endnotes}
\usepackage{rotating}
\usepackage{lscape}
\usepackage[T1]{fontenc}
\usepackage{pslatex}

\shorttitle{timining and spectral properties of X-ray dips observed from GRS 1915+105 and IGR J17091-3624}
\shortauthors{Pahari et al.}

\begin{document}

\title{Properties of unique hard X-ray dips observed from GRS 1915+105 and IGR J17091-3624 and their implications}
\author{Mayukh Pahari\altaffilmark{1,4}, J S Yadav\altaffilmark{1}, J\'{e}r$\hat{\mathrm{o}}$me Rodriguez\altaffilmark{2}, Ranjeev Misra\altaffilmark{3}, Sudip Bhattacharyya\altaffilmark{1} and S K Pandey\altaffilmark{4} }
\affil{$^1$ Tata Institute of Fundamental Research, Homi Bhabha Road, Mumbai, India; \texttt{mp@tifr.res.in\\}}
\affil{$^2$ Laboratoire AIM, CEA/IRFU - CNRS/INSU - Universit\'e Paris Diderot, CEA DSM/IRFU/SAp, F-91191 Gif-sur-Yvette, France}
\affil{$^3$ Inter University Center for Astronomy and Astrophysics, Pune, India} 
\affil{$^4$ Pt. Ravishankar Sukla University, Raipur, Chattishgarh, India}

\shorttitle{timing and spectral properties of X-ray dips observed in GRS 1915+105 and IGR J17091-3624}
\shortauthors{Pahari et al.}

\begin{abstract}

We report a comprehensive study on spectral and timing properties of hard X-ray dips uniquely observed in some so-called variability classes of the micro-quasars GRS 1915+105 and IGR J17091-3624. These dips are characterised by a sudden decline in the 2.0$-$60.0 keV X-ray intensity by a factor of 4 $-$ 12 simultaneous with the increase in hardness ratio by a factor of 2 $-$ 4. Using 31 observations of GRS 1915+105 with {\it RXTE}/PCA, we show that different behavior are observed in different types of variability classes, and we find that a dichotomy is observed between classes with abrupt transitions vs those with smoother evolution. For example, both energy-lag spectra and frequency-lag spectra of hard X-ray dips in classes with abrupt transitions and shorter dip intervals show hard-lag (hard photons lag soft photons), while both lag spectra during hard dips in classes with smoother evolution and longer dip intervals show soft-lag. Both lag time-scales are of the order of 100-600 msec. We also show that timing and spectral properties of hard X-ray dips observed in light curves of IGR J17091-3624 during its 2011 outburst are consistent with the properties of the abrupt transitions in GRS 1915+105 rather than smooth evolutions. A global correlation between the X-ray intensity cycle time and hard dip time is observed for both abrupt and smooth transition which may be due to two distinct physical processes whose time-scales are eventually correlated. We discuss implications of our results in the light of some generic models.

\end{abstract}

\keywords{accretion, accretion disks --- black hole physics  --- X-rays: individual (IGR J17091-3624, GRS 1915+105) --- X-rays: binaries}

\section {Introduction}

Rapid X-ray variabilities which lead to dramatic drops in the X-ray luminosity by a factor of $\sim$ 4 $-$ 12 in a couple of seconds and periodic repetition of such luminosity variations observed over a long time-scale, make GRS 1915+105 atypical with respect to other micro-quasars. Based on light curves and color-color diagram, the observed variabilities have been categorized into 14 different classes ($\alpha$, $\beta$, $\theta$, $\mu$, $\rho$, $\omega$, $\kappa$, $\lambda$, $\gamma$, $\delta$, $\chi$, $\phi$, $\omega$, $\zeta$; \citet{b16,b5,b99}) and broadly into three basic X-ray states $-$ (1) a low luminosity disk dominated state (state A), (2) high luminosity, thermal and/or non-thermal emission dominated state (state B) and (3) low luminosity powerlaw dominated hard state (state C). Rapid X-ray variabilities observed during $\rho$, $\alpha$, $\omega$, $\kappa$, $\lambda$, $\beta$, $\theta$ classes are thought to be due to rapid transitions among these three basic X-ray states \citep{b16,b98}. However, dissection of these classes into three basic states in \citet{b16} reveals following details : (1) among three basic states, state C is the most stable state as it has been observed for the longest time-scale in each of these classes. State C in \citet{b16} is described as a low count rate state (usually brighter than state A) where inner disk temperature is either very low ($\sim$ 0.5 keV) or there is no disk contribution. White noise variabilities observed on time scale of 1 sec. (2) Variations in occurrence time-scale of the C state is largest among three states. For example, during $\kappa$ class, state C is observed for as short as $\sim$60 sec while during $\beta$ class, it is observed for as long as $\sim$500 sec. However, it can last for few days in a non-variable class like $\chi$. (3) More importantly, variation in the luminosity of the state C is largest among three states. For example, with {\it RXTE}/PCA in the 2.0$-$60.0 keV energy band, state C is observed sometimes at 2.2 $\times$ 10$^{38}$ ergs/s during $\kappa$ class while during $\theta$ class it is sometimes observed at 11.3 $\times$ 10$^{38}$ ergs/s. 
Because of large variation in the time-scale as well as luminosity within state C, it is important to know whether all C state observations in different classes have same timing and spectral characteristics or not. If they are same, then what drives such large variations in the luminosity and time-scale within a same state ? However, no detailed analysis have been carried out to resolve these issues. 

 Significant efforts have been made to understand the origin and the nature of state C (here, we call it hard X-ray dip) in different time scales in GRS 1915+105. \citet{b64} show that hard X-ray dip spectra during $\rho$ and $\alpha$ class can be fitted with the {\tt diskbb+nthcomp} model. \citet{b63} show that both burst and hard X-ray dip spectra of $\omega$ class can be fitted using {\tt comptt+powerlaw} model. $\theta$ class spectra can be modelled using the combination of {\tt diskbb+powerlaw+comptt} where observed soft dips (following the hard dips) may be due to ejections of comptonization cloud \citep{b6}. In $\beta$, $\lambda$/$\kappa$, and $\nu$ classes, \citet{b71,b72} showed that the soft A-dips follow the longer hard C-dips which probably reflects the ejection of coronal material. 
Using phase resolved spectroscopy, \citet{b61,b60} show that heartbeat oscillations ($\rho$ class) are results of a combination of a thermal-viscous radiation pressure instability and a local Eddington limit where hard X-ray dips are resulting from the sudden radiation pressure driven evaporation. However, in a separate work, \citet{b57} show that a combination of hybrid corona and disk blackbody model can describe energy spectra of all phases in $\rho$ class including hard X-ray dips. 

Hence along-with large variations in the flux and time-scale in hard X-ray dips in different classes, there exists large degeneracy in spectral analysis \& modelling of limit cycle oscillations. One key indication from these studies is that variability properties and spectral nature of hard X-ray dips during abrupt transitions may vary from that of the hard X-ray dips during smooth transitions. This motivate us to perform detailed spectral and timing analysis of hard X-ray dips in different flux scale and time scale. 

The second motivation comes from another transient, Galactic black hole X-ray binary (BHXB) IGR J17091-3624 as it shows regular, repetitive, large amplitude oscillations in their X-ray light curve, very similar to $\rho$ cycles observed from GRS 1915+105 \citep{b26,b54,b95,b97}. Using X-ray morphology and hardness, a few more variabilities are observed from IGR J17091-3624 \citep{b81,b83}, some of which are similar to those from GRS 1915+105 and show hard X-ray dips. Time-scale and flux level of these dips are much lower than any such dip observed in GRS 1915+105. These findings greatly increase the importance of understanding the origin and nature of hard X-ray dips in these two BHXBs. 

In this work, we summarize a comprehensive study on spectral modelling and timing properties (like rms spectra, lag spectra, variabilities at different time-scales) of hard X-ray dips, frequently observed in the light curve of $\kappa$, $\lambda$, $\alpha$, $\theta$ and $\beta$ classes in GRS 1915+105 and IGR J17091-3624. We establish a global relationship between hard dip time and the cycle time in different classes of GRS 1915+105 as well as IGR J17091-3624. We show that hard dips in classes with abrupt transitions and relatively short dip time intervals (like $\kappa$, $\lambda$ classes) show hard-lag, while hard dips in classes with smooth evolutions and relatively long dip time intervals  (like $\theta$, $\beta$ and $\alpha$ classes) show soft-lag. Even lag spectra of hard dips in $\kappa$, $\lambda$ classes are different from those in $\alpha$, $\theta$ and $\beta$ classes. Although hard X-ray dips of different classes may belong to the same spectral state, i.e., state C, we detect a large variation in different spectral parameters like disk temperature, disk radius etc. A high-energy break is significantly required in the energy spectra of all high luminosity ($>$ 4 $\times$ 10$^{38}$ ergs/s) X-ray dips of $\theta$ and $\beta$ classes while low energy break/no break is required in low luminosity ($<$ 4 $\times$ 10$^{38}$ ergs/s) X-ray dips of $\kappa$, $\lambda$ and $\alpha$ classes. We show that variability properties of hard X-ray dips observed in another BHXB IGR J17091-3624 during its 2011 outburst are consistent with those of abrupt transitions rather than smooth evolutions in GRS 1915+105. We discuss our timing and spectral analysis procedure and results in section 2 and its subsections. Discussions and conclusions are provided in section 3 and its subsection.

\section{Analysis and results}

\subsection{X-ray dip selection procedure and analysis}

We analyse all {\it RXTE}/PCA observations of GRS 1915+105 between 1996 and 1998 and found 31 observations which show either of $\alpha$, $\kappa$, $\lambda$, $\beta$ and $\theta$ classes with sufficiently large number of complete X-ray dips (total dip time in each observation $>$ 500 sec). We assume that results obtained from these 31 observations are valid for all similar observations in GRS 1915+105. We exclude $\rho$ class from our analysis because hard dip intervals in this class are less than 25 sec as they are associated with high frequency oscillations and spectral parameters vary rapidly within this period. For GRS 1915+105, the full energy-channel information are only available for {\tt Standard2} data which has 16 sec time resolution. Hence energy spectral analysis and energy-dependent lag spectral analysis are not possible with these class dips. Although long, hard dip intervals have been observed several times in $\nu$ class, within 2 years of observational data considered in our work (including observations mentioned in \citet{b16,b94}), we find only one observation where the complete cycle of $\nu$ class (complete dip interval+complete burst interval) is observed (MJD 50381.49; Obs-ID : 10408-01-44-00). All other observations show either incomplete dip intervals or partial burst intervals. Hence, we are able to perform partial spectral and timing analysis during $\nu$ class. From complete data of 2011 outburst in IGR J17091-3624, we find 6 {\it RXTE}/PCA observations which show hard X-ray dips similar to GRS 1915+105. Due to low count rate (order of magnitude lower than GRS 1915+105) and small hard dip intervals ($\sim$ 20 sec), we merge all hard dip intervals from 6 observations to obtain a statistically significant signal strength and reasonably long dip interval ($\sim$ 154.8 sec) for further spectral and timing analysis. Observation details are provided in Table 1. 

\begin{table}
 \centering
 \caption{Hard X-ray dip observation details and energy spectral parameters along with 1$\sigma$ errorbars fitted with {\tt diskbb+bknpower} and {\tt diskbb+powerlaw} model (shown by stars) for different classes in GRS 1915+105 and IGR J17091-3624.}
\begin{center}
\scalebox{0.50}{%
\begin{tabular}{ccccccccccccccc}
\hline
Obs. date & Class  & Total dip time & kT$_{in}$ & R$_{in}$ & PL index & break energy & PL index & L$_{total}$ & F$_{total}$ & F$_{diskbb}$ & F$_{bknpower}$/ & $\chi^2/dof$ & $\chi^2/dof$ \\
(in MJD)     & & (sec) & (keV)       & (km) & $\Gamma_1,{\tt bknpower}$/ & (keV) & ($\Gamma_2,{\tt bknpower}$) &  &  & & F$_{powerlaw}$ & ($\tt bknpower$) & ($\tt powerlaw$)  \\  
& & & & & $\Gamma_{\tt powerlaw}$ & & & & & & & & \\
\hline
$^*$50575.93 & $\alpha$ & 4300.3 & $0.87^{+0.03}_{-0.02}$ & $99^{+11}_{-9}$ & $2.26^{+0.02}_{-0.01}$ & -- & -- & 3.41 & 1.98$^{+0.11}_{-0.09}$ & 0.38$^{+0.03}_{-0.05}$ & 1.59$^{+0.12}_{-0.11}$ & 40.6/48 & 46.1/50 \\
$^*$50577.56 & $\alpha$ & 3403.8 & $0.88^{+0.04}_{-0.03}$ & $97^{+9}_{-10}$ & $2.25^{+0.03}_{-0.02}$ & -- & -- & 3.56 & 2.07$^{+0.09}_{-0.08}$ & 0.37$^{+0.04}_{-0.06}$ & 1.69$^{+0.15}_{-0.10}$ & 41.1/48 & 45.7/50 \\
$^*$50586.63 & $\alpha$ & 3754.7 & $0.84^{+0.05}_{-0.04}$ & $109^{+12}_{-11}$ & $2.23^{+0.02}_{-0.01}$ & -- & -- & 3.49 & 2.03$^{+0.13}_{-0.07}$ & 0.40$^{+0.05}_{-0.06}$ & 1.62$^{+0.10}_{-0.13}$ & 43.2/48 & 49.1/50 \\
$^*$50567.74 & $\alpha$ & 2503.24 & $0.95^{+0.06}_{-0.04}$ & $80^{+8}_{-11}$ & $2.31^{+0.04}_{-0.05}$ & -- & -- & 3.69 & 2.15$^{+0.12}_{-0.08}$ & 0.43$^{+0.04}_{-0.02}$ & 1.70$^{+0.11}_{-0.07}$ & 42.2/48 & 43.3/50 \\
50363.27 & $\lambda$  & 1161.5 & $1.16^{+0.03}_{-0.02}$ & $82^{+9}_{-11}$ & $2.86^{+0.04}_{-0.04}$ & $10.6^{+0.1}_{-0.2}$ & $2.39^{+0.02}_{-0.03}$ & 3.97 & 2.31$^{+0.16}_{-0.21}$ & 0.41$^{+0.04}_{-0.06}$ & 1.90$^{+0.14}_{-0.12}$ & $39.3/48$ & 68.6/50 \\
50639.63  & $\lambda$  & 1117.6 & $1.18^{+0.03}_{-0.03}$ & $84^{+8}_{-9}$ & $2.87^{+0.03}_{-0.04}$ & $10.5^{+0.1}_{-0.1}$ & $2.46^{+0.03}_{-0.03}$ & 3.38 & 1.96$^{+0.19}_{-0.13}$ & 0.26$^{+0.02}_{-0.03}$ & 1.68$^{+0.12}_{-0.09}$ & $46.2/48$ & 88.9/50 \\
50641.21 & $\lambda$  & 876.2 & $0.75^{+0.03}_{-0.02}$ & $80^{+10}_{-8}$ & $2.91^{+0.04}_{-0.05}$ & $10.8^{+0.9}_{-0.6}$ & $2.21^{+0.04}_{-0.03}$ & 3.93 & 2.29$^{+0.5}_{-0.34}$ & 0.42$^{+0.08}_{-0.05}$ & 1.86$^{+0.12}_{-0.38}$ & $48.2/48$ & 125.8/50 \\
50641.79 & $\lambda$  & 895.4 & $0.68^{+0.04}_{-0.05}$ & $69^{+7}_{-8}$ & $2.92^{+0.05}_{-0.04}$ & $10.9^{+0.8}_{-0.4}$ & $2.19^{+0.04}_{-0.03}$ & 3.91 & 2.28$^{+0.16}_{-0.41}$ & 0.41$^{+0.06}_{-0.06}$ & 1.85$^{+0.12}_{-0.11}$ & $45.6/48$ & 81.8/50 \\
50627.55 & $\kappa$  & 2725.5 & $0.82^{+0.02}_{-0.02}$ & $139^{+12}_{-14}$ & $2.87^{+0.04}_{-0.04}$ & $10.8^{+0.1}_{-0.1}$ & $2.19^{+0.01}_{-0.01}$ & 2.43 & 1.41$^{+0.09}_{-0.13}$ & 0.19$^{+0.02}_{-0.02}$ & 1.22$^{+0.08}_{-0.10}$ & $42.3/48$ & 113.3/50 \\
50627.92 & $\kappa$ & 2225.8 & $0.87^{+0.04}_{-0.05}$ & $68^{+8}_{-10}$ & $2.90^{+0.07}_{-0.08}$ & $11.1^{+0.3}_{-0.2}$ & $ 2.17^{+0.01}_{-0.01}$ & 2.82 & 1.64$^{+0.19}_{-0.15}$ & 0.15$^{+0.01}_{-0.02}$ & 1.45$^{+0.09}_{-0.13}$ & $37.4/48$ & 77.0/50 \\  
51284.16 & $\kappa$ & 535.6 & $0.93^{+0.03}_{-0.02}$ & $89^{+8}_{-11}$ & $2.48^{+0.04}_{-0.06}$ & $12.9^{+0.2}_{-0.1}$ & $2.66^{+0.09}_{-0.05}$ & 2.56 & 1.49$^{+0.02}_{-0.03}$ & 0.18$^{+0.01}_{-0.02}$ & 1.26$^{+0.09}_{-0.11}$ &$39.8/48$ & 66.3/50  \\
51284.09 & $\kappa$ & 772.2 & $1.13^{+0.03}_{-0.04}$ & $70^{+8}_{-10}$ & $2.64^{+0.03}_{-0.04}$ & $11.7^{+0.1}_{-0.1}$ & $2.39^{+0.03}_{-0.04}$ & 3.41 & 1.98$^{+0.09}_{-0.14}$ & 0.57$^{+0.05}_{-0.07}$ & 1.40$^{+0.11}_{-0.12}$ & $44.2/48$ & 71.3/50 \\
$^*$51284.28 & $\kappa$  & 1651.5 & $1.35^{+0.06}_{-0.05}$ & $58^{+7}_{-8}$ & $2.82^{+0.08}_{-0.07}$ & -- & -- & 3.89 & 2.27$^{+0.11}_{-0.09}$ & 0.41$^{+0.03}_{-0.02}$ & 1.84$^{+0.18}_{-0.11}$ & $33.3/48$ & 46.5/50\\
51284.35 & $\kappa$  & 1661.6 & $1.16^{+0.05}_{-0.03}$ & $94^{+8}_{-12}$ & $2.75^{+0.06}_{-0.05}$ & $11.5^{+0.2}_{-0.2}$ & $2.49^{+0.04}_{-0.05}$ & 3.54 & 2.06$^{+0.19}_{-0.12}$ & 0.55$^{+0.04}_{-0.01}$ & 1.50$^{+0.14}_{-0.09}$ & $50.5/48$ & 68.7/50 \\
$^*$50617.61 & $\kappa$  & 864.5 & $1.25^{+0.05}_{-0.03}$ & $77^{+8}_{-6}$ & $2.78^{+0.04}_{-0.02}$ & -- & -- & 2.72 & 1.58$^{+0.12}_{-0.14}$ & 0.22$^{+0.04}_{-0.03}$ & 1.31$^{+0.06}_{-0.09}$ & $38.3/48$ & 46.6/50 \\
$^*$50617.54 & $\kappa$ & 2330.7 & $1.30^{+0.04}_{-0.02}$ & $51^{+5}_{-9}$ & $2.79^{+0.04}_{-0.03}$ & -- & -- & 2.37 & 1.38$^{+0.19}_{-0.09}$ & 0.37$^{+0.03}_{-0.04}$ & 0.99$^{+0.09}_{-0.11}$ & $28.9/48$ & 38.8/50 \\ 
50688.67 & $\beta$ & 912.3 & $0.97^{+0.02}_{-0.02}$ & $211^{+20}_{-13}$ & $2.38^{+0.02}_{-0.02}$ & $18.2^{+1.1}_{-1.0}$ & $2.68^{+0.04}_{-0.03}$ & 4.28 & 2.49$^{+0.19}_{-0.16}$ & 0.68$^{+0.05}_{-0.02}$ & 1.79$^{+0.15}_{-0.10}$ &$49.6/48$ & 105.6/50 \\
50691.38 & $\beta$ & 2011.7 & $0.98^{+0.02}_{-0.02}$ & $207^{+17}_{-15}$ & $2.46^{+0.04}_{-0.03}$ & $18.9^{+0.9}_{-1.0}$ & $2.90^{+0.05}_{-0.03}$ & 4.69 & 2.73$^{+0.21}_{-0.14}$ & 0.82$^{+0.02}_{-0.08}$ & 1.90$^{+0.21}_{-0.16}$ & $52.5/48$ & 249.9/50  \\
50698.56 & $\beta$ & 1015.3 & $1.14^{+0.03}_{-0.03}$ & $116^{+7}_{-5}$ & $2.66^{+0.03}_{-0.04}$ & $16.6^{+0.9}_{-1.0}$ & $2.86^{+0.03}_{-0.03}$ & 4.63 & 2.69$^{+0.11}_{-0.21}$ & 0.76$^{+0.04}_{-0.01}$ & 1.92$^{+0.15}_{-0.11}$ & $46.4/48$ & 88.8/50  \\
50700.91 & $\beta$ & 925.9 & $1.07^{+0.03}_{-0.03}$ & $122^{+7}_{-10}$ & $2.46^{+0.04}_{-0.03}$ & $17.9^{+0.9}_{-1.0}$ & $2.92^{+0.05}_{-0.03}$ & 4.45 & 2.59$^{+0.09}_{-0.13}$ & 0.93$^{+0.09}_{-0.02}$ & 1.59$^{+0.16}_{-0.10}$ & $42.5/48$ & 176.6/50 \\
50800.23 & $\beta$ & 1020.4 & $1.03^{+0.03}_{-0.03}$ & $131^{+9}_{-11}$ & $2.44^{+0.03}_{-0.02}$ & $18.3^{+1.1}_{-1.0}$ & $2.89^{+0.05}_{-0.03}$ & 4.33 & 2.52$^{+0.09}_{-0.16}$ & 0.69$^{+0.03}_{-0.03}$ & 1.82$^{+0.19}_{-0.11}$ & $40.9/48$ & 105.7/50 \\
50802.35 & $\beta$ & 2210.12 & $1.20^{+0.05}_{-0.04}$ & $82^{+12}_{-7}$ & $2.66^{+0.04}_{-0.05}$ & $17.6^{+1.0}_{-0.9}$ & $3.03^{+0.04}_{-0.05}$ & 4.52 & 2.63$^{+0.21}_{-0.26}$ & 0.77$^{+0.08}_{-0.04}$ & 1.84$^{+0.15}_{-0.10}$ &$51.3/48$ & 244.7/50 \\
50674.41 & $\beta$ & 2784.22 & $0.99^{+0.02}_{-0.02}$ & $219^{+20}_{-14}$ & $2.32^{+0.04}_{-0.03}$ & $18.6^{+0.9}_{-0.9}$ &  $2.56^{+0.06}_{-0.05}$ & 4.61 & 2.68$^{+0.19}_{-0.12}$ & 0.77$^{+0.02}_{-0.03}$ & 1.89$^{+0.12}_{-0.16}$ & $47.1/48$ & 70.5/50 \\
50675.78 & $\beta$ & 965.7 & $0.96^{+0.02}_{-0.02}$ & $198^{+16}_{-11}$ & $2.33^{+0.03}_{-0.02}$ & $19.1^{+1.1}_{-1.0}$ & $2.57^{+0.05}_{-0.04}$ & 4.16 & 2.42$^{+0.18}_{-0.14}$ & 0.92$^{+0.09}_{-0.03}$ & 1.49$^{+0.16}_{-0.09}$ & $40.7/48$ & 81.4/50 \\
50706.17 & $\theta$ & 1008.67 & $1.42^{+0.01}_{-0.02}$ & $72^{+5}_{-3}$ & $2.71^{+0.04}_{-0.05}$ & $16.4^{+1.0}_{-0.9}$ & $3.24^{+0.07}_{-0.08}$ & 7.88 & 4.58$^{+0.13}_{-0.18}$ & 0.77$^{+0.06}_{-0.08}$ & 3.81$^{+0.15}_{-0.08}$ & $46.9/48$ & 143.5/50 \\
50706.96 & $\theta$ & 1920.43 & $1.48^{+0.02}_{-0.03}$ & $85^{+5}_{-6}$ & $2.61^{+0.06}_{-0.05}$ & $15.9^{+1.1}_{-0.9}$ & $3.30^{+0.07}_{-0.08}$ & 9.19 & 5.34$^{+0.23}_{-0.22}$ & 1.28$^{+0.08}_{-0.04}$ & 4.06$^{+0.16}_{-0.11}$ & $38.1/48$ & 282.4/50 \\
50707.23 & $\theta$ & 800.21 & $1.54^{+0.03}_{-0.01}$ & $84^{+4}_{-5}$ & $2.71^{+0.05}_{-0.04}$ & $16.1^{+1.0}_{-0.9}$ & $3.24^{+0.07}_{-0.05}$ & 11.61 & 6.75$^{+0.31}_{-0.28}$ & 1.54$^{+0.09}_{-0.12}$ & 5.15$^{+0.15}_{-0.11}$ & $39.6/48$ & 159.2/50   \\
50707.88 & $\theta$ & 720.14 & $1.52^{+0.03}_{-0.02}$ & $86^{+5}_{-6}$ & $2.70^{+0.05}_{-0.06}$ & $15.9^{+1.0}_{-0.75}$ & $3.37^{+0.08}_{-0.06}$ & 11.28 & 6.56$^{+0.23}_{-0.33}$ & 1.47$^{+0.12}_{-0.11}$ & 5.04$^{+0.18}_{-0.06}$ & $42.3/48$ & 245.5/50 \\
50250.16 & $\theta$ & 752.17 & $1.45^{+0.02}_{-0.03}$ & $92^{+6}_{-6}$ & $2.64^{+0.04}_{-0.03}$ & $16.8^{+0.6}_{-0.8}$ & $3.33^{+0.06}_{-0.05}$ & 9.86 & 5.73$^{+0.20}_{-0.33}$ & 1.22$^{+0.19}_{-0.11}$ & 4.45$^{+0.11}_{-0.13}$ & $47.4/48$ & 80.1/50\\
50250.91 & $\theta$ & 928.55 & $1.36^{+0.03}_{-0.02}$ & $102^{+4}_{-4}$ & $2.67^{+0.05}_{-0.07}$ & $15.9^{+1.0}_{-0.9}$ & $3.29^{+0.09}_{-0.07}$ & 8.86 & 5.15$^{+0.29}_{-0.21}$ & 1.21$^{+0.11}_{-0.13}$ & 3.94$^{+0.13}_{-0.23}$ & $49.5/48$ & 76.6/50\\
50696.37 & $\theta$ & 1248.65 & $1.51^{+0.03}_{-0.04}$ & $77^{+5}_{-5}$ & $2.77^{+0.04}_{-0.05}$ & $16.1^{+1.0}_{-0.9}$ & $3.34^{+0.07}_{-0.06}$ & 11.78 & 6.85$^{+0.48}_{-0.31}$ & 1.19$^{+0.11}_{-0.09}$ & 5.54$^{+0.14}_{-0.05}$ & $46.6/48$ & 215.2/50  \\
\hline
-- & IGR dips & 154.8 & $0.54^{+0.31}_{-0.09}$ & -- & $3.36^{+0.58}_{-1.12}$ & $11.8^{+2.1}_{-1.0}$ & $1.53^{+0.55}_{-0.07}$ & -- & $0.049^{+0.015}_{-0.003}$ & $0.005^{+0.003}_{-0.001}$ & $0.044^{+0.010}_{-0.008}$ & 47.3/48 & 78.1/50 \\
\hline
\end{tabular}}
\end{center}
\tablecomments {kT$_{in}$ is the inner disk temperature in keV, R$_{in}$ is the apparent inner disk radius in km (calculated for GRS 1915+105 assuming the distance of 12$\pm$2 kpc, disk inclination of 70$^o$ and color correction factor of 1.7), $\Gamma_1,{\tt bknpower}$ and $\Gamma_2,{\tt bknpower}$ are power law indices of {\tt bknpower} model, $\Gamma_{powerlaw}$ is the power law index of {\tt powerlaw} model, F$_{total}$, F$_{diskbb}$, F$_{bknpower}$ and F$_{powerlaw}$ are total unabsorbed flux, unabsorbed flux due to {\tt diskbb}, {\tt bknpower} and {\tt powerlaw} model in the 2.0$-$60.0 keV energy range respectively. L$_{total}$ is the total luminosity in the unit of 10$^{38}$ ergs/s. The unit of flux is 10$^{-8}$ ergs/s/cm$^{-2}$.}
\end{table}

For each observation, we determine the dip time from each dip profile (steady, persistent flux level within two consecutive rapid state transitions; similar to \citet{b94}) considering both the X-ray intensity profile and hardness ratio (HR) profile. We consider the significant change in the HR values comparing to the flaring interval as well as the X-ray count rate close to the minimum. This allow us to exclude mixing the X-ray peaks ending the X-ray dips, even if they are hard. For example, in $\beta$ class, we found that during dip-to-burst transition, the HR below 0.08 corresponds to the steep rise phase of the X-ray spike following the X-ray dip in the intensity profile (middle right panel of Figure 1) and this steep rise cannot be the part of dip intervals. Thus, time intervals in the $\beta$ class for which HR is below 0.08, do not belong to the dip intervals as observed in X-ray intensity profile. The reason here is that spectral parameters describing the disk blackbody and comptonization are already significantly different than during the lower part of the hard X-ray dips \citep{b72}. Considering these facts, we choose dip time intervals when HR is approximately $\ge$ 0.1 for $\kappa$ and $\lambda$ classes, $\ge$ 0.14 for $\alpha$ class, $\ge$ 0.08 for $\beta$ class, $\ge$ 0.07 for $\theta$ class and $\ge$ 0.49 for IGR J17091-3624. Although small dispersions have been observed around selected HR limits in $\theta$ and $\alpha$ class dips, choices of these HR values in different classes ensure that the dip intervals in different classes do not get mixed up with the sharp-decay and the sharp-rise part of the intensity profiles and appear to be reasonable and consistent for all observations. HR is defined as the ratio of background subtracted count rate between 12.0-60.0 keV and 2.0-12.0 keV.
Selected intervals are shown in red in each panel of Figure 1 for different classes in GRS 1915+105 and IGR J17091-3624. In all panels of Figure 1, light curves where minimum count rate is normalized to 1 count/sec and corresponding time is normalized to 0 sec are shown at the top and their hardness ratio are shown at the bottom for $\kappa$ (top left panel), $\lambda$ (middle left panel), IGR J17091-3624 (bottom left), $\alpha$ (top right panel), $\beta$ (middle right panel) and $\theta$ (bottom right panel) classes respectively. In order to avoid dispersion due to small time bin-size (1 sec), we compute the count rate minimum position using large bin-size (10 sec) and count rate minimum position from both bin-sizes are found consistent. In all panels, examples of dip time intervals are shown by vertical black, dotted lines and HR limits for selecting dip time intervals are shown by horizontal dotted lines. It is interesting to note that dips in the X-ray intensity during $\theta$ class (bottom right panel of Figure 1), is actually soft (i.e., low HR values) rather than hard. This behavior is opposite to X-ray dips in all other classes. Hence hard time intervals in the $\theta$ class have higher intensity than actual X-ray dips. We select high intensity hard time intervals in the $\theta$ class as persistent dip time for further analysis.

\begin{figure*}
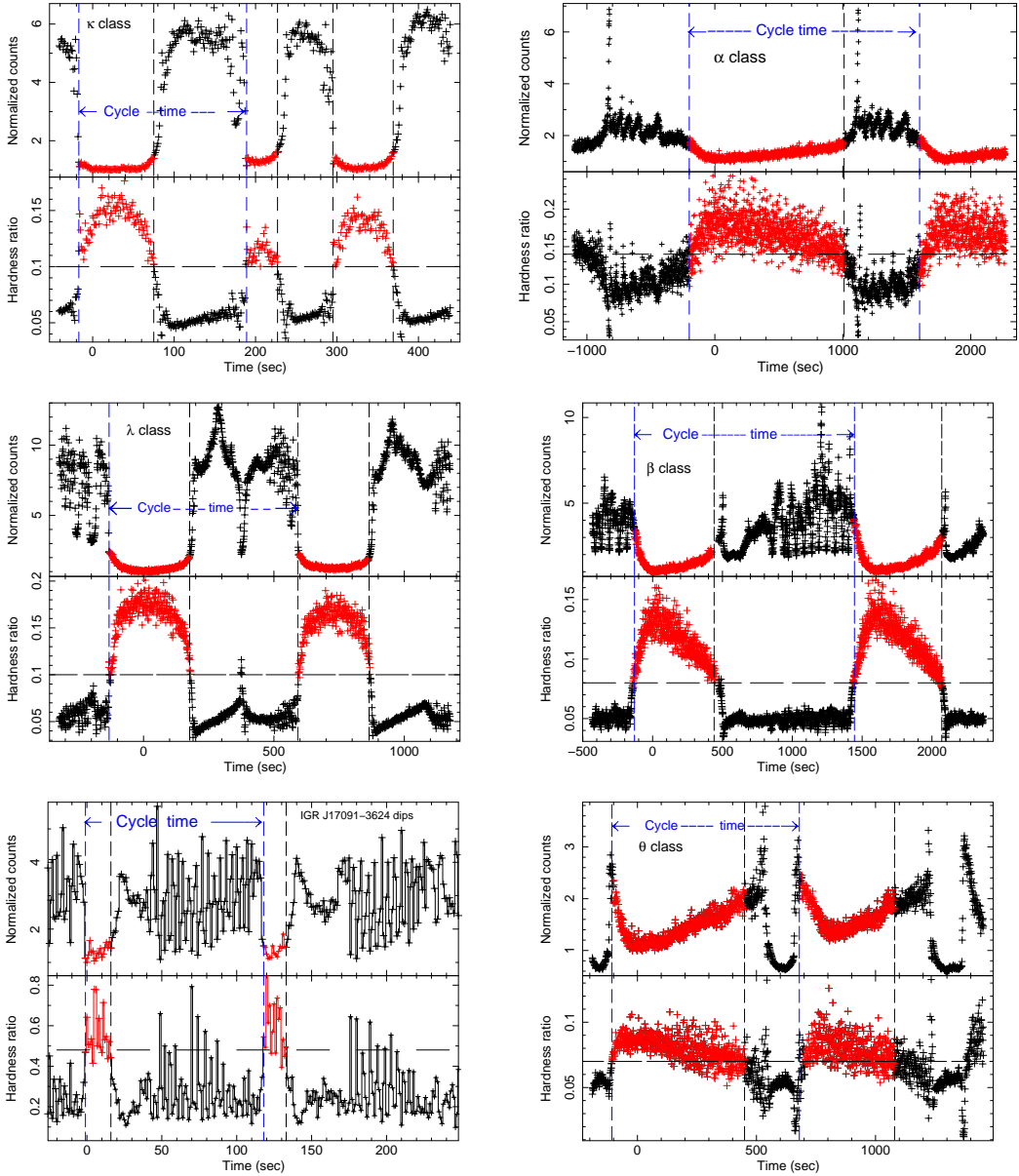

\centering
\begin{tabular}{cc}
\includegraphics*[width=2.0in,angle=-90]{fig1a.ps} &
\includegraphics*[width=2.0in,angle=-90]{fig1b.ps}\\
\includegraphics*[width=2.0in,angle=-90]{fig1c.ps} &
\includegraphics*[width=2.0in,angle=-90]{fig1d.ps}\\
\includegraphics*[width=2.0in,angle=-90]{fig1e1.ps} &
\includegraphics*[width=2.0in,angle=-90]{fig1f.ps}\\
\end{tabular}
\caption{One sec binned, normalized {\it RXTE}/PCA X-ray intensity profile (where the minimum of the count rate is normalized to 1 count/sec and corresponds to 0 sec on time axis) and simultaneous hardness ratio profile during $\kappa$ class (top left panel), $\lambda$ class (middle left panel), $\alpha$ class (top right panel), $\beta$ class (middle right panel) and $\theta$ class (bottom right panel) in GRS 1915+105 and in IGR J17091-3624 (bottom left panel). In all panels, selection of hard dip intervals are shown in red.}
\label{period-mjd}
\end{figure*}

\subsection{Correlation among different time-scales}

\subsubsection{X-ray dip time vs. total cycle time}

Careful observation reveals that time-scales in dip intervals are different within a class as well different classes. To quantify it, we plot the hard dip+dip-burst transition+following burst (i.e., total cycle) time-scales as a function of the hard dip time-scales (left panel of Figure 2). We show example selections of cycle time-scales in each panel of Figure 1 using blue, vertical dotted lines. It is observed that occurrence of X-ray dips in IGR J17091-3624 is extremely irregular. For both sources, we define the cycle as a precise pattern seen to repeat always in the same fashion. In IGR J17091-3624 intensity profiles, X-ray dips are always followed by a short spike and a rapid oscillations (see the selection marked as ``cycle time" in the bottom left panel of Figure 1). This is similar to the pattern observed in GRS 1915+105.

\begin{figure}
\centering
\begin{tabular}{cc}
\includegraphics*[width=3.2in]{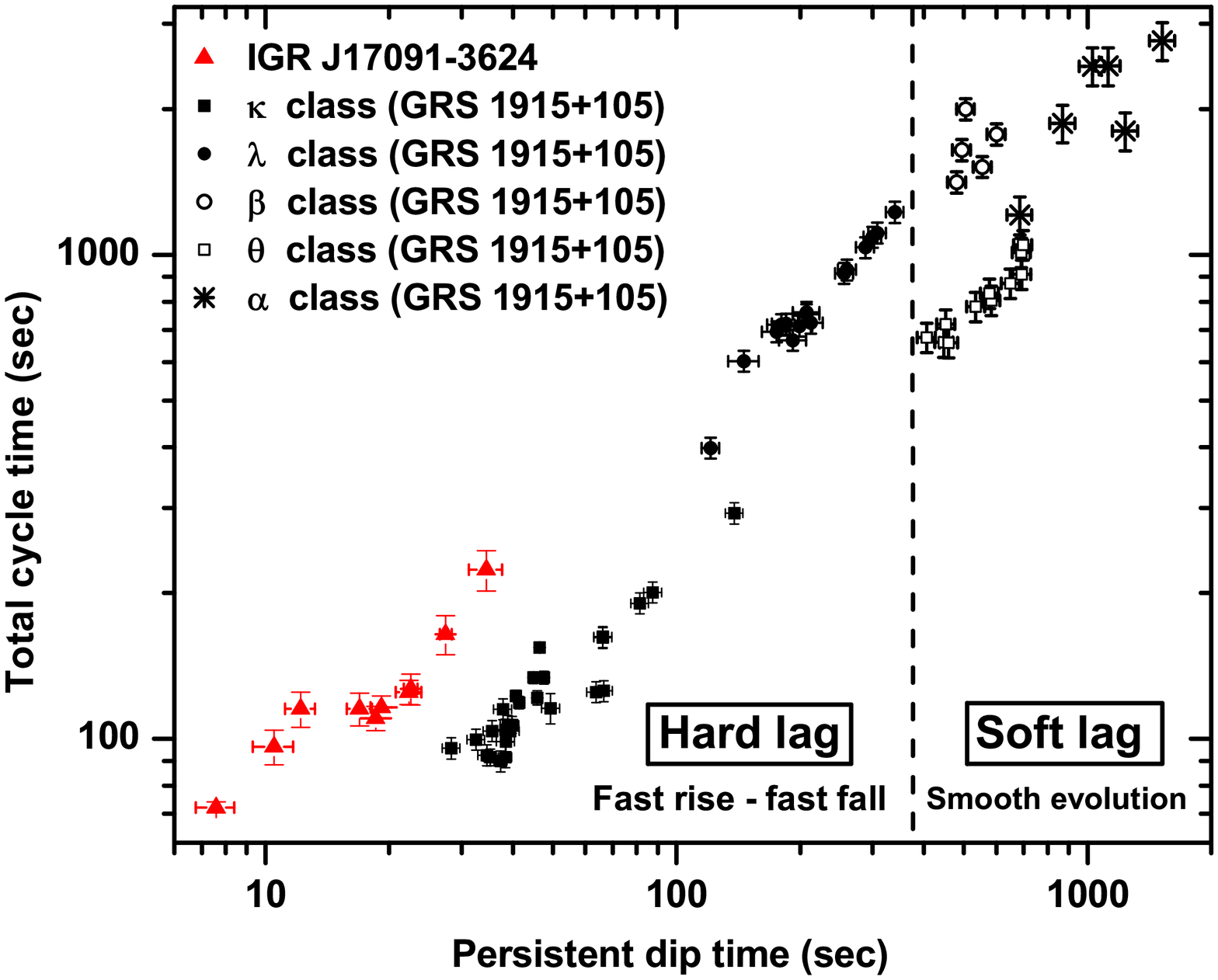} &
\includegraphics*[width=3.2in]{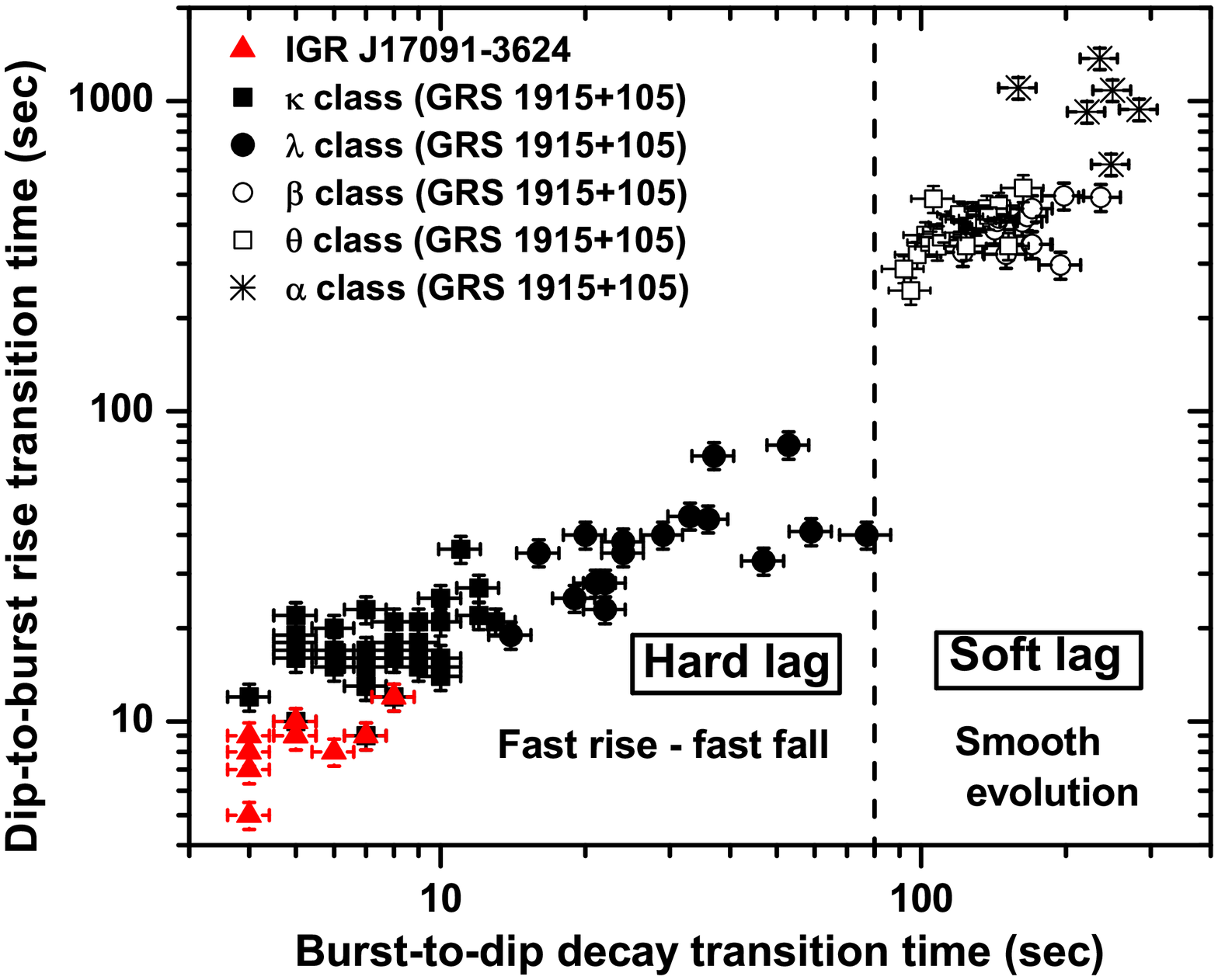} \\
\end{tabular}
\caption{{\it Left panel :} Plot of total cycle time of each oscillation in different classes of GRS 1915+105 and IGR J17091-3624 as a function of persistent hard dip time. {\it Right panel :} Plot of dip-to-burst transition time as a function of burst-to-dip transition time during different classes in GRS 1915+105 and IGR J17091-3624. A strong global correlation is observed in different time-scales.}
\label{period-mjd}
\end{figure}

Left panel of Figure 2 shows that there exist a strong global correlation between the total cycle time and the persistent dip time in different classes from both sources. Previously, using the $\kappa$ class observations which show abrupt transitions during X-ray dips, it is shown that there exists a correlation between dip time and burst time \citep{b10,b2}. In our work, along with the $\kappa$ class, we include other X-ray dips from GRS 1915+105 where abrupt transitions are observed like the $\lambda$ class and X-ray dips observed in another BHXB IGR J17091-3624. In addition to these, we also include classes where smooth evolutions are observed like $\theta$, $\beta$ and $\alpha$ classes. They show complex burst structures like X-ray spikes, soft dip, rapid oscillations etc. which are entirely different from the steady, simple and near-Eddington burst structures observed during  $\kappa$ or $\lambda$ classes. In spite of differences in burst structures in different classes, we show that a highly significant correlation exists among different time-scales in different classes. Thus our correlation is global as it incorporates X-ray dips and cycle time-scales from different classes as well as different sources.

We fit the plot of two time-scales with a constant, linear and quadratic function. Both the linear and the quadratic functions fit the data with the significance of $>$9$\sigma$ and $\sim$ 6$\sigma$ (using F-test) over the constant respectively. We prefer to choose linear function as it has one less parameter than the quadratic function. We quantify the correlation by measuring the Pearson product-moment correlation coefficient (PMCC)\citep{b50}. Due to the different HR behavior observed in the $\theta$ class, we measure the PMCC twice $-$ with and without the $\theta$ class. the PMCC comes out to be 0.906 when X-ray dip and cycle time from the $\theta$ class are included and the PMCC comes out to be 0.948 when X-ray dip and cycle time from the $\theta$ class are excluded. Thus exclusion of the $\theta$ class improves the correlation measurement. It may be noted that the average dip time intervals of IGR J17091-3624 occupy the lowest values in the parameter space and it is slightly above the trend in the parameter space observed in GRS 1915+105.

\subsubsection{burst-to-dip transition time vs. dip-to-burst transition time}

Another important difference, as observed from Figure 1, is that $\kappa$ and $\lambda$ classes have fast-fall and fast-rise type dip structures (average slope of fall profile \& rise profile are $\sim$0.25 $-$ 0.29 normalized count rate/sec and $\sim$0.31 $-$ 0.36 normalized count rate/sec respectively) while $\theta$, $\beta$ and $\alpha$ classes have slow-fall and slow-rise type dip structures  (average slope of fall profile \& rise profile are $\sim$0.018 $-$ 0.034 normalized count rate/sec, $\sim$0.004 $-$ 0.013 normalized count rate/sec and $\sim$0.001 $-$ 0.003 normalized count rate/sec respectively). From all panels of Figure 1 it is seen that the X-ray intensity profile shape of dip intervals during $\kappa$, $\lambda$ classes are different from that during $\beta$, $\theta$ and $\alpha$ classes. Thus we define burst-to-dip and dip-to-burst transition time-scales in different manner for both group of classes. During each cycle of the light curve profile in $\kappa$/$\lambda$ classes, we calculate the average burst count rate from the burst profile and average dip count rate from the persistent dip profile. Then these time-scales are measured as the time taken by the source when its intensity just departs the average burst count rate and reach the last moment just before the average dip count rate and vice-versa. For $\beta$, $\theta$ and $\alpha$ classes, the burst-to-dip fall time is defined as the time to reach the minimum of the dip profile from the peak of the last burst before the dip and the dip-to-burst rise time is defined as the time to reach from the minimum of the dip profile to the peak of the first burst after the dip. Although X-ray dips in IGR J17091-3624 also show fast-fall and fast-rise type structures, qualitatively similar to that of the $\kappa$ class, due to existance of rapid oscillations in burst structures, both time scales are measured using second method. When we plot the dip-to-burst transition time as a function of burst-to-dip transition time (right panel of Figure 2), we find that (1) this two time-scales are eventually different for different classes which is consistent with the prediction of the disk instability model \citep{b7}. (2) IGR J17091-3624 shows fastest transitions among all classes from both sources and transition time scales in IGR J17091-3624, $\kappa$, $\lambda$ classes differ significantly from that of $\theta$, $\beta$ and $\alpha$ classes. We calculate the ratio of transition time scales to dip time scales in different classes. We found that for all observations of $\beta$, $\theta$ and $\alpha$ classes, average values of the ratio are 0.93 $\pm$ 0.11, 0.83 $\pm$ 0.12 and 0.71 $\pm$0.14 respectively while for all observations of $\kappa$, $\lambda$ classes and IGR J17091-3624 dips, average values of the ratio are 0.25 $\pm$ 0.07, 0.37 $\pm$ 0.04 and 0.49 $\pm$ 0.19 respectively.

\begin{figure}
\centering
\begin{tabular}{cc}
\includegraphics*[width=3.2in]{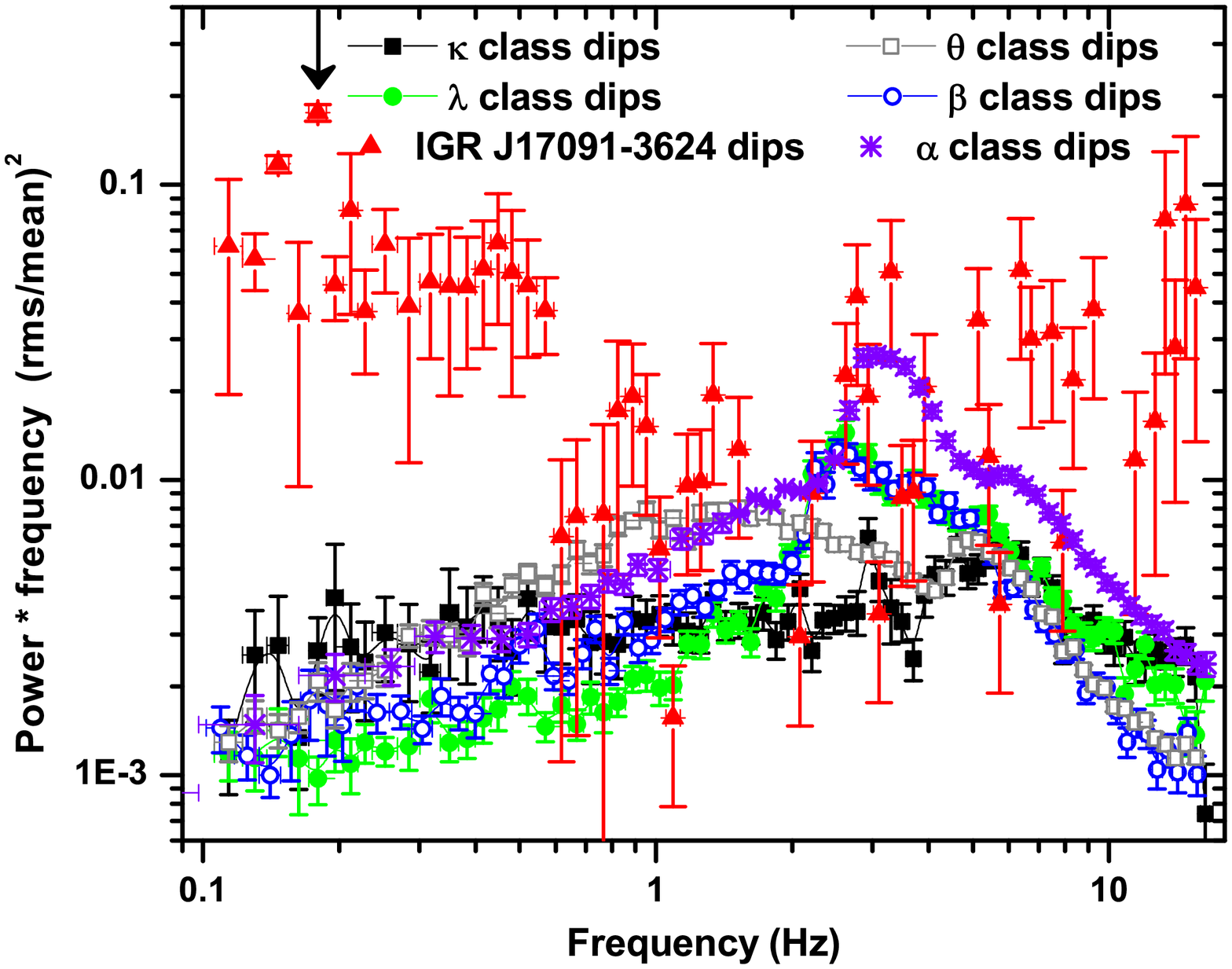} &
\includegraphics*[width=3.2in]{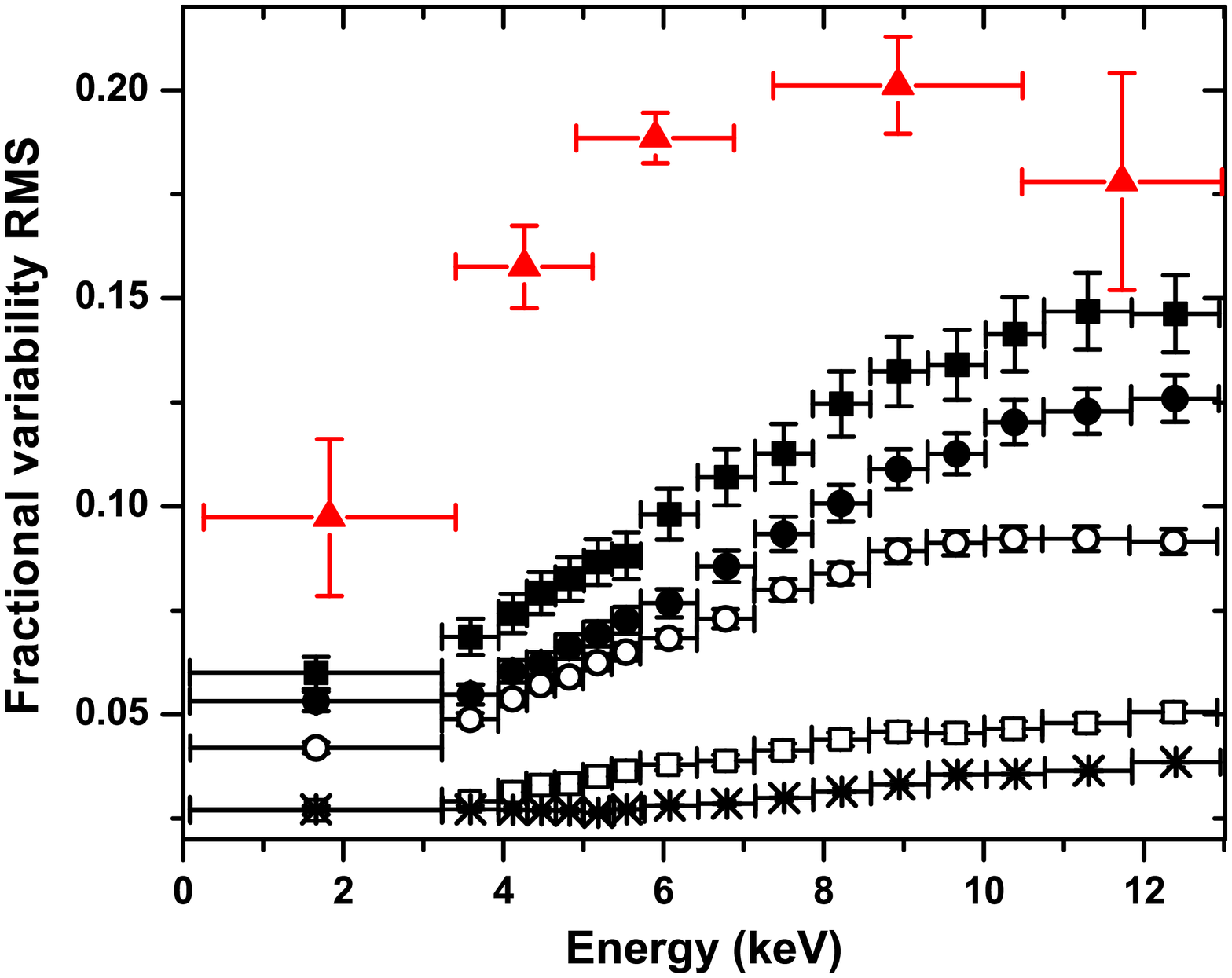} \\
\end{tabular}
\caption{{\it Left panel :} Plot of rms normalized and white noise subtracted power density spectra of hard X-ray dips during $\kappa$ (black solid squares), $\lambda$ (green solid circles), $\beta$ (blue solid squares), $\theta$ (grey solid squares) and $\alpha$ (violet stars) classes in GRS 1915+105 and in IGR J17091-3624 (red triangles). The arrow at 0.18 Hz denotes the frequency at which lag and coherence are calculated. {\it Right panel :} Plot of intrinsic fractional variability rms spectra as a function of photon energy are shown during hard X-ray dips in $\kappa$ (black solid squares), $\lambda$ (black solid circles), $\beta$ (black hollow circles), $\theta$ (black hollow squares) and $\alpha$ (black crosses) classes in GRS 1915+105 and in IGR J17091-3624 (red triangles).}
\label{period-mjd}
\end{figure}

\subsection {Timing analysis and results}

For dip intervals during $\kappa$, $\lambda$, $\beta$, $\theta$ and $\alpha$ classes in GRS 1915+105 and in IGR J17091-3624, we produce white noise subtracted and rms normalized power density spectra (PDS; in the unit of (rms/mean)$^2$) in the frequency range 0.1 $-$ 20 Hz. Left panel of Figure 3 shows PDS for different X-ray dip intervals. Results from IGR J17091-3624 are shown by red triangles, in all panels of Figure 2, 3, 4 and left panel of Figure 5. To study the nature of variability during X-ray dips, we perform energy dependent study of Poisson fluctuation subtracted fractional variability rms spectra (right panel of Figure 3), during hard X-ray dips. From the rms spectra it may be noted that at all energy bands, fractional variability rms is very high during X-ray dips in IGR J17091-3624 compared to X-ray dips of all classes in GRS 1915+105. 
Even if the Poisson noise is higher, simply because the source is fainter, the source variability is significant and thus well estimated. The values of the rms even with large errors, are still significantly higher than those of GRS 1915+105 ($>$ 3$\sigma$ in the  5-7 keV range for example; see right panel of Figure 3).

In order to calculate time-lag and coherence in different energy bands, we develop and use a Fortran code which uses the cross spectrum technique to calculate the phase-lag and time-lag between two time series. To develop the code, we follow formulations and assumptions thoroughly given in \citet{b52}. Within error-bars, our results agree well with results from another code {\tt GHATS 1.0.2} (T. Belloni; private communications). Energy dependent lag spectra and coherence are calculated at the Fourier frequency of 0.18 Hz as IGR J17091-3624 show highest power at this frequency (left panel of Figure 3). Photon energy dependent lag spectra, coherence and Fourier frequency dependent lag spectra of hard X-ray dips are shown in the top left, top right and bottom left panels of Figure 4 for different classes of GRS 1915+105 and IGR J17091-3624 respectively. While calculating time-lag and coherence in different energy bands, we use 3.2$-$4.2 keV and 0.3$-$3.4 keV as the reference band in GRS 1915+105 and IGR J17091-3624 respectively. The reason for choosing different energy bands as reference bands is that in all energy bands, the X-ray intensity of IGR J17091-3624 is the order of magnitude lower than GRS 1915+105. We observe that a narrow energy band (for example, 3.2-4.2 keV which is used here as a reference band) in GRS 1915+105 has sufficiently high signal-to-noise ratio. While to get the same signal-to-noise ratio in IGR J17091-3624, a wider energy band is necessary. Thus to ensure best possible result from both sources we make such choices of energy ranges for reference bands. Additionally, both sources during X-ray dips are powerlaw dominated ($>$ 80\% of total flux), hence we assume that choice of different reference band, even below 5 keV, do not mix different physical processes.  

From top panels and bottom left panel of Figure 4, we find that for all observations, energy-dependent and frequency-dependent time-lag spectra show hard-lag (hard photons lag soft photons) during $\kappa$ and $\lambda$ classes while it show soft-lag during $\beta$, $\theta$ and $\alpha$ classes consistently. X-ray dips during $\nu$ class show soft-lag where timing properties and lag time-scales are similar to that of the $\beta$ class. However, frequency dependent time-lag spectra show very complex behavior. For different classes, we calculate average time delay for different classes (average of time delays measured from different observations) as a function of average persistent dip time which is shown in the bottom right panel of Figure 4. Here we measure average time delay between 10.4-12.8 keV and 0.3-3.4 keV for both GRS 1915+105 and IGR J17091-3624. Average time delay values in different classes as well as trends in lag observed in this plot are consistent with the energy-lag spectra (top left panel of Figure 4). This confirms that the choice of different reference bands in computing energy-lag spectra do not affect results. In a log-linear scale (Y(x) = A* ln(x) + B), the plot can be fitted by a linear function with the slope of -0.27 $\pm$ 0.05. A constant can be ruled out with the significance of 4.4$\sigma$. IGR J17091-3624 show hard-lag during X-ray dips similar to that observed during $\kappa$ class. In all cases, the time-lag increases and the coherence decreases at higher energies.

\begin{figure}
\centering
\begin{tabular}{cc}
\includegraphics*[width=3.2in]{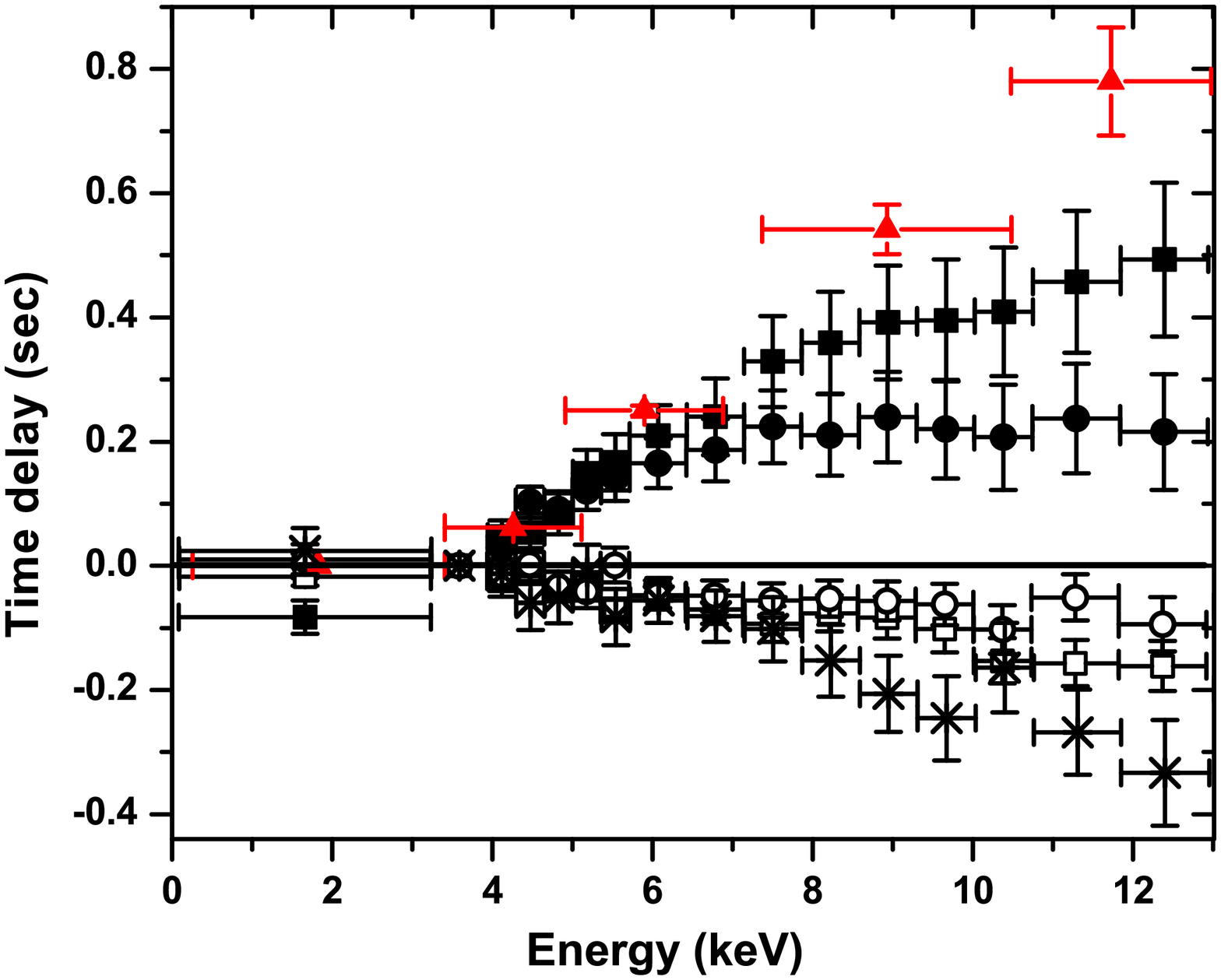} &
\includegraphics*[width=3.2in]{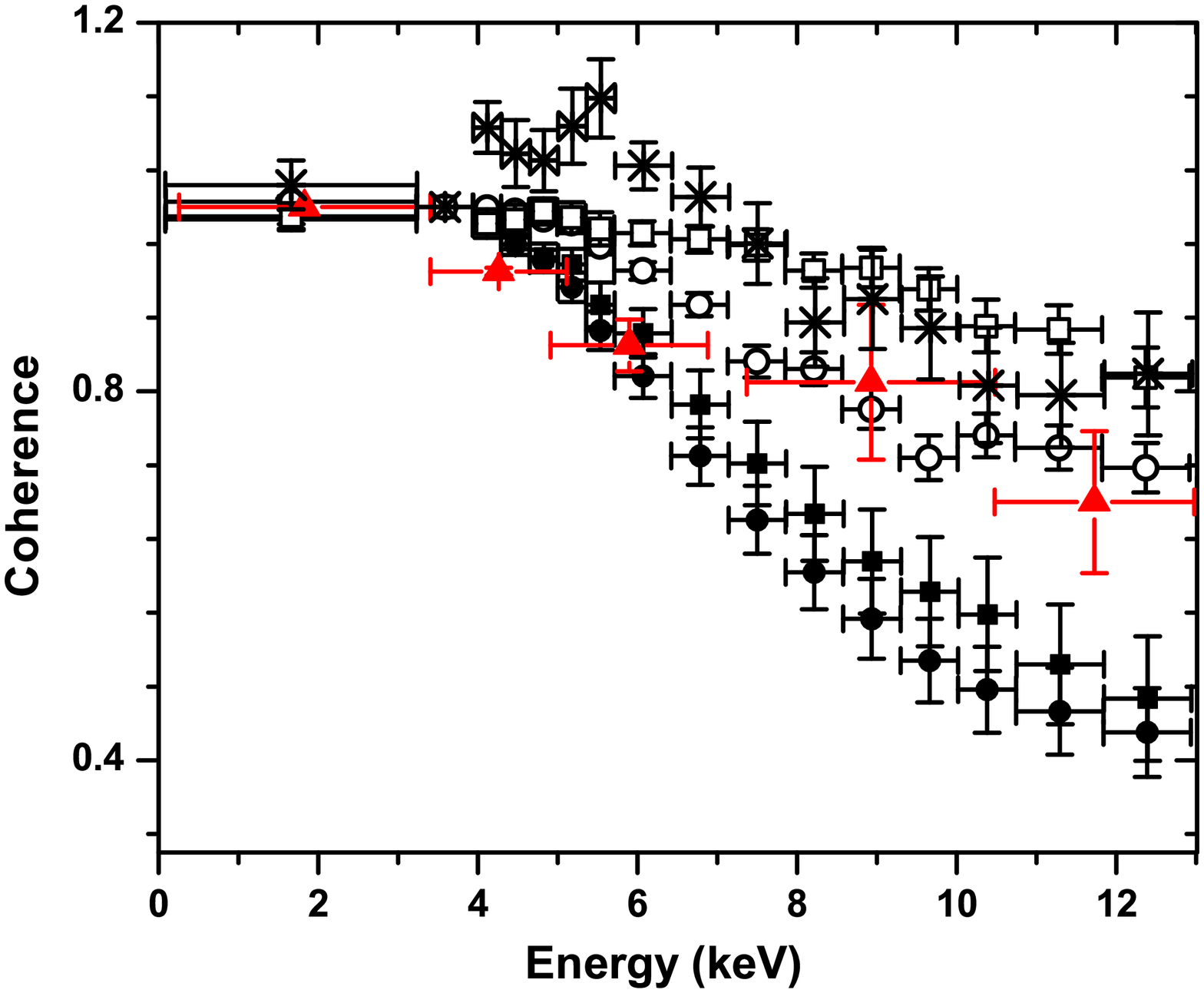}  \\
\includegraphics*[width=3.2in]{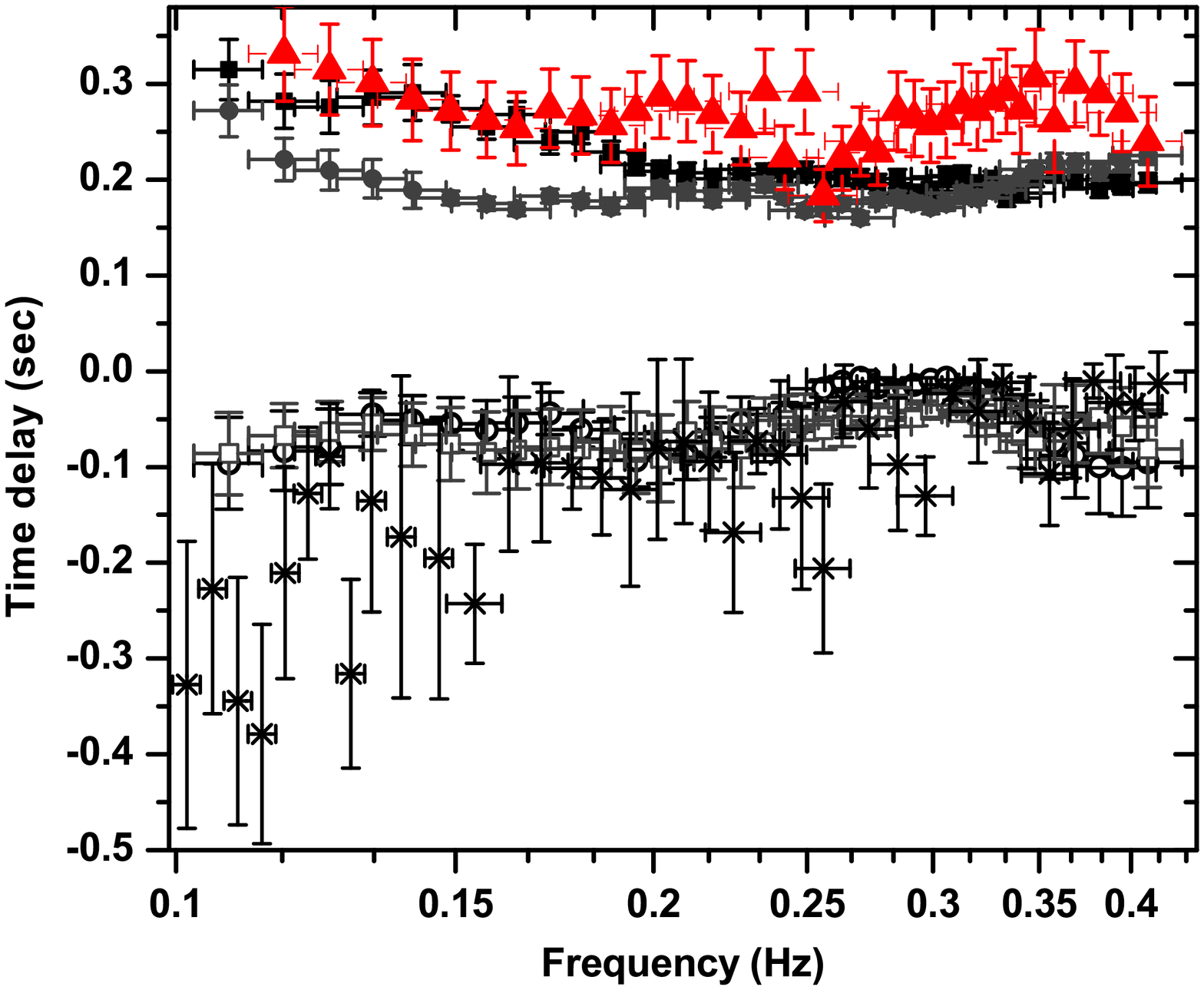}  &
\includegraphics*[width=3.2in]{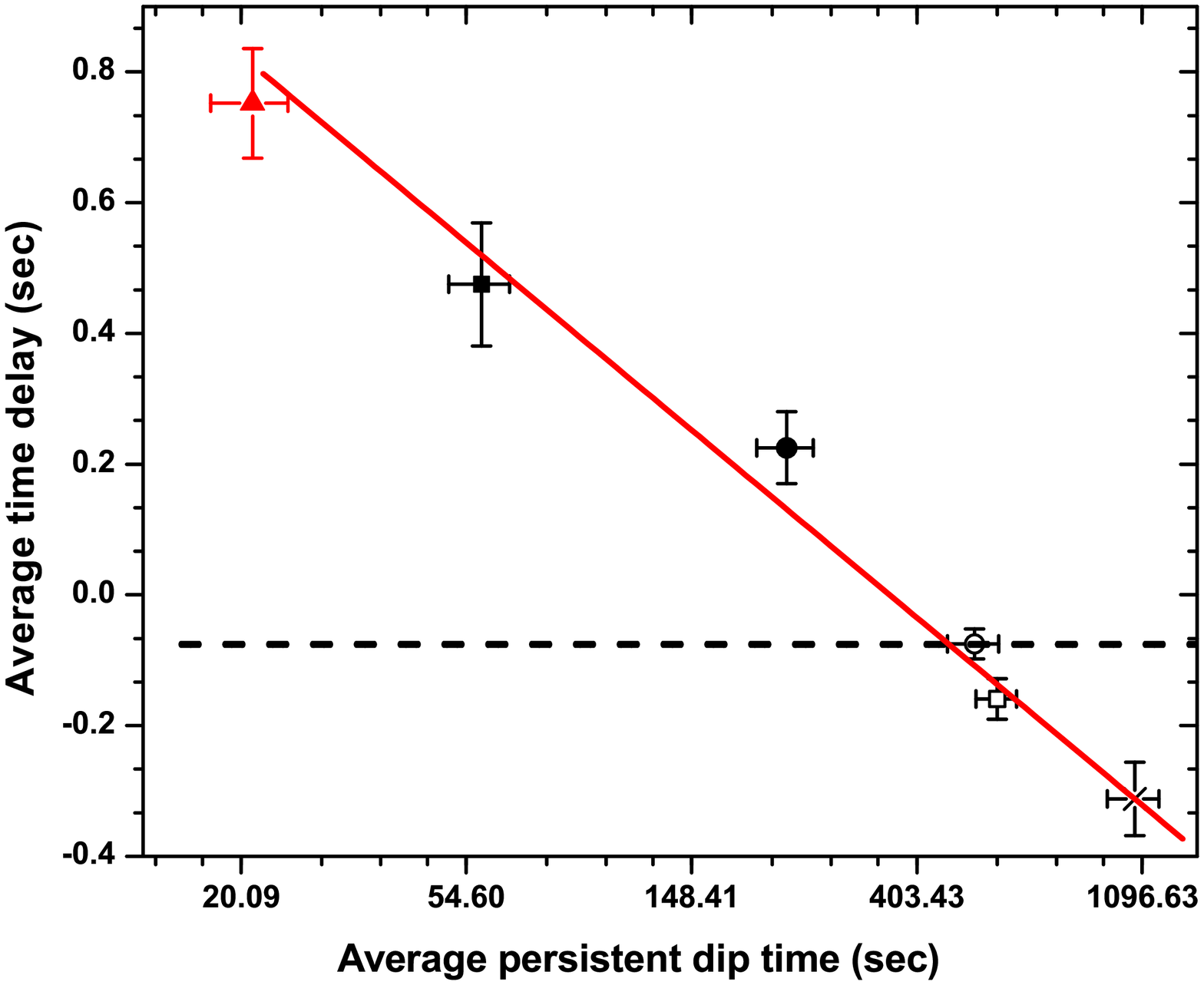}  \\
\end{tabular}
\caption{{\it Top left \& top right panel :} Plot of time-lag spectra and coherence function as a function of photon energy, {\it bottom left panel :} time-lag spectra as a function of Fourier frequency and {\it Bottom right panel :} average time delay as a function of average persistent hard X-ray dip time are shown during hard X-ray dips in $\kappa$ (black solid squares), $\lambda$ (black solid circles), $\beta$ (black hollow circles), $\theta$ (black hollow squares) and $\alpha$ (black crosses) classes in GRS 1915+105 and in IGR J17091-3624 (red triangles). The log-linear plot of average time delay vs average X-ray dip time can be fitted with a straight line (shown by red dotted line) with the slope of -0.27 $\pm$ 0.05.}
\label{period-mjd}
\end{figure}

\subsection{Spectral analysis and results}

We extract dip spectra using {\tt Standard2} data with 16 sec time resolution. For each observation, we merge time intervals corresponding to dip profiles and use them to extract spectra. 
To fit spectra, we try several models ({\tt diskbb}, {\tt powerlaw}, {\tt bknpower}, {\tt comptt} in {\tt XSpec} and combination of them). We use 3.0$-$25.0 keV energy range for spectral analysis and assume 1\% systematic error while fitting \citep{b73}. We find that except {\tt diskbb+bknpower} and {\tt diskbb+powerlaw} model, all other models fit the dip spectra with unacceptably high reduced $\chi^2$ ($>$ 1.2) for all classes. For all observations of the $\alpha$ class and three observations of the $\kappa$ class, both models provide similar and acceptable reduced $\chi^2$ (see Table 1). For these observations, we prefer {\tt diskbb+powerlaw} model as it has two less parameters than the {\tt diskbb+bknpower} model. For other observations, even with 1\% systematic errors, very high reduced $\chi^2$ are observed with the {\tt diskbb+powerlaw} model. This implies that this model is not appropriate.

Observing a break in spectra, we use {\tt diskbb+bknpower} model which provides acceptable reduced $\chi^2$. Fitted parameter values along with 1$\sigma$ error-bars, reduced $\chi^2$ values for both {\tt diskbb+bknpower} and {\tt diskbb+powerlaw} models are provided in Table 1. We add a Gaussian line to all spetra in GRS 1915+105 fixed at 6.3 keV. We use neutral hydrogen column density fixed at 6 $\times$ 10$^{22}$ cm$^{-2}$ for GRS 1915+105 \citep{b9} and 0.91 $\times$ 10$^{22}$ cm$^{-2}$ for IGR J17091-3624 \citep{b97}. From the Table 1, it may be noted that a high-energy break in powerlaw continuum is significantly detected in all high luminosity ($>$ 4 $\times$ 10$^{38}$ ergs/s) X-ray dips of $\theta$ and $\beta$ classes while low energy break/no break has been detected in all low luminosity ($<$ 4 $\times$ 10$^{38}$ ergs/s) X-ray dips of $\kappa$, $\lambda$ and $\alpha$ classes. X-ray dips of the $\nu$ class show luminosity and energy spectral properties similar to the $\beta$ class where spectra energy break in powerlaw continuum is detected significantly at 17.34$^{+1.35}_{-2.43}$ keV. The plot of disk temperature as a function of total unabsorbed flux (left panel of Figure 5) shows that disk temperature increases at higher total flux. 
A large variation in apparent inner disk radius ($\sim$ 52 km $-$ 221 km) is observed when plotted against total flux although no correlation is found (right panel of Figure 5). Flux-temperature relationship measured with hard X-ray dip spectra in IGR J17091-3624 is shown by red triangle in the left panel of Figure 5 where the flux of IGR J17091-3624 is multiplied by 20 for easy comparison.  

\begin{figure}[t]
\begin{tabular}{cc}
\includegraphics*[scale=0.3]{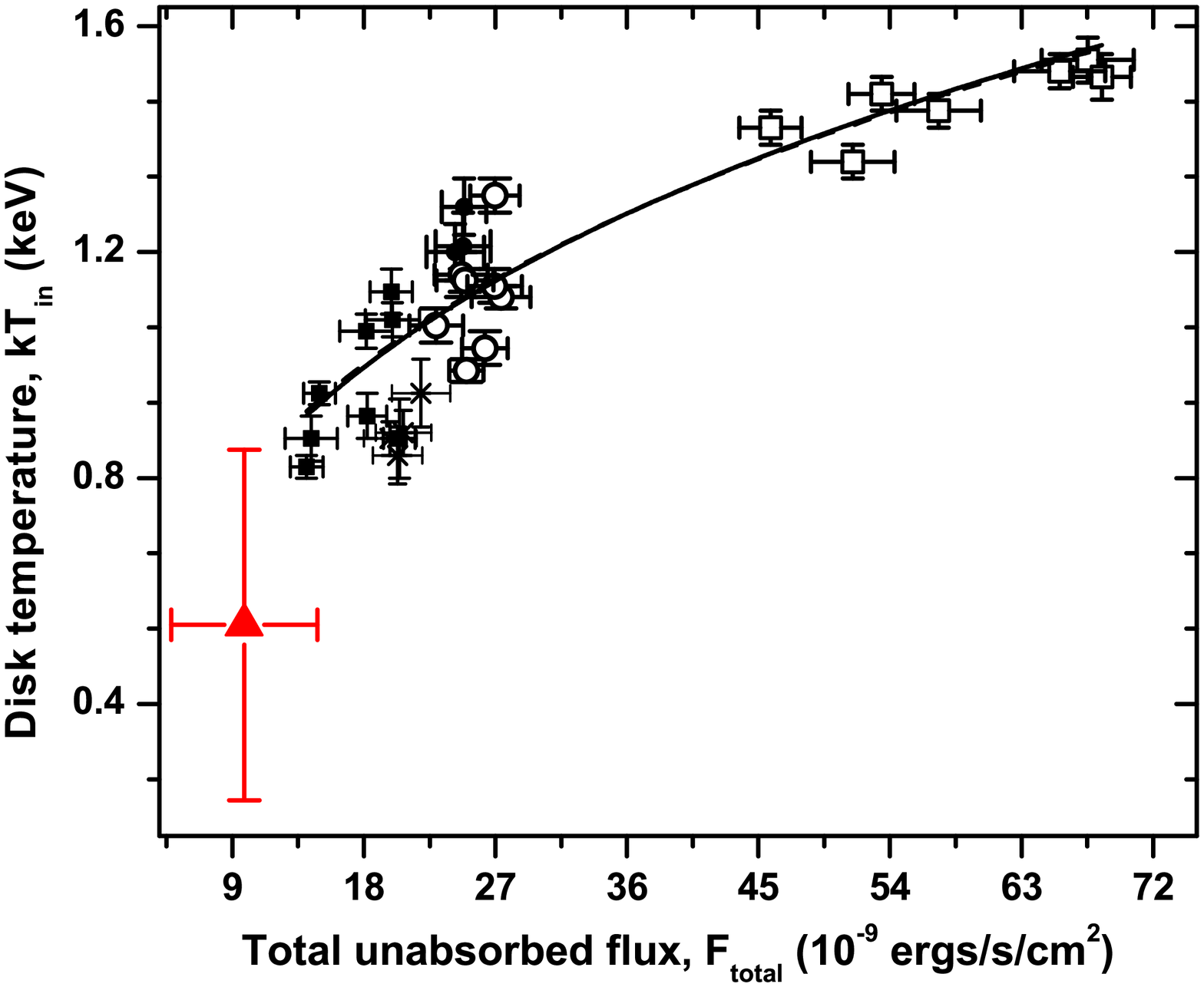}  &
\includegraphics*[scale=0.3]{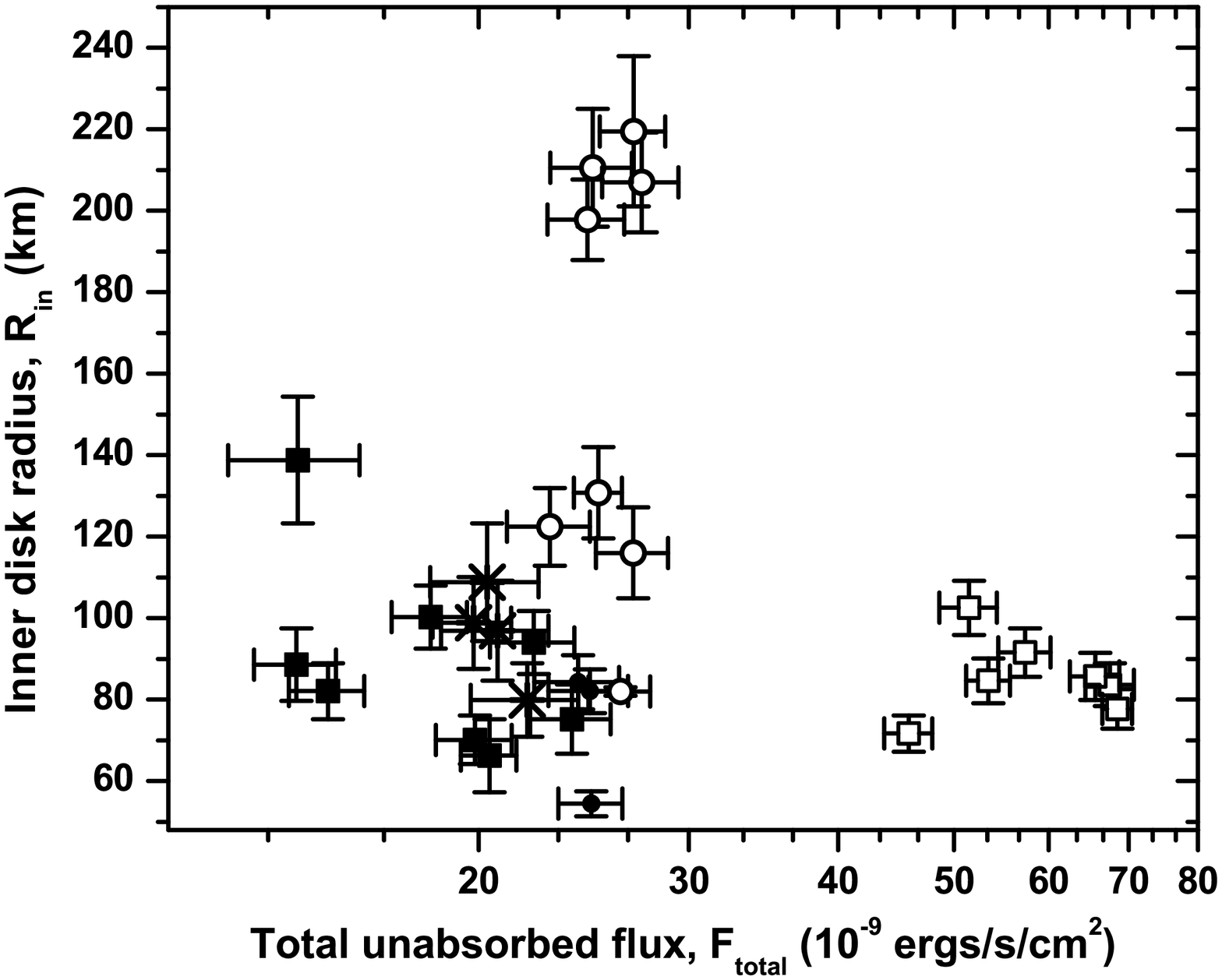}  \\
\end{tabular}
\caption{Plot of inner disk temperature (left panel) and inner disk radius (right panel) as a function of total unabsorbed flux in 2.0-60.0 keV energy range. In these panels, all symbols have same definition as used in Figure 3. Total flux $-$ temperature relationship in GRS 1915+105 can be fitted by a standard blackbody emission (i.e., F$_{total}$ $\propto$ T$^4$; shown by solid line). Flux-temperature value during hard dip of IGR J17091-3624 shown by a red triangle in the left panel where flux is multiplied by 20.}
\label{period-mjd}
\end{figure}

\section{Discussion and conclusions}

In this work, using both timing and spectral properties of hard X-ray dips observed from GRS 1915+105, we show for the first time that hard X-ray dips with the abrupt transition (like in $\kappa$, $\lambda$ classes) show hard-lag while hard X-ray dips with the smooth evolution (like in $\beta$, $\theta$, $\alpha$ classes) show soft-lag. Frequency dependent lag spectral shape during hard-lag is different from that during soft-lag. Moreover, a strong global correlation between cycle time and hard dip time for different classes is observed. 

The robustness and integrity of results from GRS 1915+105 increase when we add results from the analysis of hard X-ray dips from another BHXB IGR J17091-3624.  
Hard X-ray dips from this source occupy extreme position in the timing parameter space. For example, comparing both sources using Figure 2,3,4 and 5, we find the following: (1) From the HR diagram, X-ray dips in IGR J17091-3624 are found to be hardest among all dips, (2) Average persistent dip time, average dip to burst transition time as well as average cycle time are found to be shortest in IGR J17091-3624, (3) intrinsic fractional variability rms in IGR J17091-3624 is highest at all energy bands measured up to 14 keV, (4) time delays are longest in IGR J17091-3624 above 8 keV and (5) disk temperature is lowest among all. 

The correlation between the total length of a cycle
and the time spent in the hard dip (Figure 2) clearly shows that
a fundamental mechanism common to both sources, but also to all variability classes
(although they obviously have differences as illustrated by the difference in
the time-lag) is at the origin of the variability patterns. 
 Previously, the relationship between the X-ray dip time and the following burst time has been established using thermal viscous instability model where the total time of the cycle is set by the time scale of the instability, while the dip/burst time-scales are set by the shape of the wave in mass accretion rate (for example, see \citet{b3,b7}). Since the dip duration is the function of mass accretion rate, then the instability model can predicts larger Radio ejection events during $\beta$ and $\theta$ classes which are associated with X-ray dips longer than that of $\kappa$ and $\lambda$ classes. However, this scenario may not be true in the $\alpha$ class where very long X-ray dips (longer than $\beta$ and $\theta$ class, see Figure 2) are associated with relatively low accretion rate and fainter Radio ejection events than the $\beta$ class (see Figure 7 of \citet{b94} and \citet{b84}). Because of a possible inconsistency between observed dip intervals in the $\omega$ class and dip intervals calculated using model-predicted relationship between inner disk radius and dip time-scales \citep{b85}, the connection of these two time-scales and ejection event amplitudes, remains to be clearly understood. 

The correlation between two time-scales observed in Figure 2 needs to be explored further and could basically have three possible origins: (1) it is the length of the hard dip intervals that eventually sets the total time-scale of a given cycle, (2) during dip time intervals and during cyclic pattern/burst structures, there may exists two distinct processes whose time-scales are correlated and (3) the total time of a cycle influences on all its individual sub-structures and thus the length of the hard dip intervals.
The first possibility can be ruled out on the basis of the following argument. From Figure 4, it may be noted that $\theta$, $\beta$ and $\alpha$ classes show soft-lag during dip intervals. If the observed soft-lag in these classes is due to a single physical process which is common to all, then, following the first possibility, it is difficult to explain why the process that is responsible for large oscillations during non-dip time intervals in $\beta$ or $\alpha$ classes, is the same process which produces soft dips during non-dip time intervals in the $\theta$ class. It is even more difficult to comprehend as the average dip time during $\beta$ and $\theta$ classes are nearly equal (see Figure 2). These facts strengthen the second possibility that the dip behavior and the non-dip behavior of all classes may be due to two distinct physical processes whose time-scales are eventually correlated. However, observing Figure 2, it may also be possible that there exists a critical cycle time above which a rapid oscillation in burst structure (as observed in $\beta$ and $\alpha$ classes) is observed. If the observed cycle time is below the critical value, it can show smooth structures during bursts (like in $\kappa$ or $\lambda$ classes)
or soft dips (as observed in the $\theta$ class). However, the spectral nature of bursts may be very different from that of soft dips and hence lag properties. Because of the fact that the average cycle time in the $\lambda$ class is similar to that of the $\theta$ class within error-bars and given that we have few measurements, it is difficult to completely rule out the third possibility. We can not include results from hard dips in IGR J17091-3624 into this discussion as both time intervals of different classes in IGR J17091-3624 (if similar classes are present in this source \citep{b83}) may be scaled down by some factors. 
Interestingly, in GRS 1915+105, a powerlaw dependent connection between the length of hard dips, in all the
different classes (but $\rho$) and the amplitude of the subsequent ejection
is also known \citep{b71,b94}. Additionally, this may indicate that
the cycle length are in-fine set by the amount of material accumulated and later
ejected. In this case, however, the differences in the time scales of the spectral variations
and the associated differences of time-lag we discover in our study (Sec. 2.2 \& 2.3) remain
to be understood.

\subsection{Implications of lag behavior}

Our results show that all hard X-ray dips may not have the same physical characteristics and
hence a simple single scenario such as the repetitive evacuation and refilling of inner accretion disk \citep{b7}, may not be applicable to all the different oscillations observed in GRS 1915+105 (see discussions of \citet{b85}). Soft and hard-lags observed during different oscillations do not seem to have a straight forward explanation. This has also been discussed greatly in \citet{b74}. Reflection from the disk or reverberation lags has been invoked
to explain both hard and soft-lags for Active Galactic Nuclei \citep{b70, b53, b69}. Depending upon the plasma optical depth and temperature gradient in the Comptonizing region, \citet{b92} proposed two different mechanisms of Compton scattering which lead to Comptonization delays that may also give rise to hard and soft-lags. The central idea of this model is that if the bulk comptonization or the corona with a temperature gradient has small optical depth, then it will show hard-lag as the scattering fraction is less. On the other hand, corona with large velocity gradient and high optical depth can show soft-lag as the compton upscattered photons, originated inside the corona, can release their energy to the relatively cooler, low-velocity electrons at the outer surface of the corona. This is indeed a versatile model, since depending on the optical depth and the temperature gradient with radii, one can get either hard or soft-lags. Similarly, soft-lag due to comptonization delays is also explained by a Compton up-scattering model \citep{b93}, where the oscillation in plasma temperature is responded by the variation in the Wien blackbody temperature of the soft seed photons. However, the magnitude of Comptonization lags is of the order of $\mu$S, i.e., of the order of light crossing time-scales. Hence these models are applicable to those systems where the observed lag is small like lags at high QPO frequencies in neutron star X-ray binaries. In the present case, the observed lags are of the order of $\sim$ 100-600 msecs which would correspond to light travel-time distance of thousands of Schwarzschild radii. Hence, it is unlikely that the observed lag in GRS 1915+105 and IGR J17091-3624 are due to reflection or Comptonization lags. A hard-lag, observed in the time-scale of $\sim$ 1 sec during heartbeat oscillations in GRS 1915+105, is explained using time-dependent computations of thermal-viscous evolution of an accretion disk which is coupled to a corona by mass exchange \citep{b88}. However, the calculation gives no clue whether this scenario can be applied to the soft-lag, observed in the order of milliseconds, during hard X-ray dips. 

Previously, \citet{b7,b10} noticed that the hard X-ray dip part of the cycle is associated with the viscous time-scale of the radiation-pressure supported inner disk. From our analysis, we find that hard X-ray dips longer than $\sim$400 sec show soft-lag while hard X-ray dips shorter than $\sim$400 sec show hard-lag where both lags at different viscous time-scales can be fitted by a straight line in the log-linear scale (bottom right panel of Figure 4). A natural implication of this result is that lag time-scales and sign as well, are strongly connected to the viscous time-scales.  Hence, the observation of lag time-scales much longer than the light crossing ones can be explained by the viscous propagation model. Here viscous fluctuations originating at an outer radii propagate inwards causing accretion rate fluctuations in the inner parts. Not only can the model explain the overall power spectra seen in black hole binaries \citep{b91}, but also explain the frequency dependent time-lags seen in these sources \citep{b89, b90, b67, b74}. However, since the outer region of the disk typically emits lower energy photons than the inner ones, the expected time-lag from this model should be hard. If the soft-lag, observed in some of the X-ray dips, is to be interpreted in the viscous fluctuation model, then the physical state of the system is unusual with high energy photons coming from outer regions. Hence, the viscous propagation model is unable to explain soft-lag. It may also be speculated that different spectral components react to the basic underlying parameters like accretion rate, viscous stress, magnetic fields etc after some different time-delays, giving rise to varied time-lag behavior in photon energies. At least phenomenological (if not fundamental) based models are needed to develop in order to explain the observed time-lag which should be able to differentiate between any spectrally degenerate models. Long term, multi-wavelength monitoring of both sources by future missions like ASTROSAT may also provide crucial information.   

\section{Acknowledgements}

We are thankful to the anonymous referee for providing crucial comments/suggestions which improves the paper. MP is thankful to D. Altamirano for his comments and suggestions. This research has made use of the General High-energy Aperiodic Timing Software (GHATS) package developed by T.M. Belloni at INAF - Osservatorio Astronomico di Brera and data obtained through the High Energy Astrophysics Science Archive Research Center online service, provided by the NASA/Goddard Space Flight Center.

\clearpage

\label{lastpage}

\end{document}